\documentclass[aip,dcolumn,reprint,amsmath,amssymb]{revtex4-1}

\usepackage{graphicx}
\usepackage{dcolumn}
\usepackage{bm}

\usepackage[colorlinks]{hyperref}
\hypersetup{allcolors=[RGB]{0,0,255}}

\usepackage[version=3]{mhchem} 
\usepackage{siunitx}

\begin{document}







\preprint{Applied Physics Letters, \textbf{111}, 113106 \,(2017)}

\title{High-speed XYZ-nanopositioner for scanning ion conductance microscopy} 



\author{Shinji Watanabe}
\email[]{wshinji@se.kanazawa-u.ac.jp}
\thanks{}
\affiliation{Bio-AFM Frontier Research Center, Institute of Science and Engineering, Kanazawa University, Kakuma-machi, Kanazawa 920-1192, Japan}

\author{Toshio Ando}
\email[]{tando@staff.kanazawa-u.ac.jp}
\thanks{}
\affiliation{Bio-AFM Frontier Research Center, Institute of Science and Engineering, Kanazawa University, Kakuma-machi, Kanazawa 920-1192, Japan}



\begin{abstract}

We describe a tip-scan-type high-speed XYZ-nanopositioner designed for scanning ion conductance microscopy (SICM). The nanopipette probe is mounted in the center of a hollow piezoactuator, both ends of which are attached to identical diaphragm flexures, for Z-positioning. This design minimizes the generation of undesirable mechanical vibrations. Mechanical amplification is used to increase the XY-travel range of the nanopositioner. The first resonance frequencies of the nanopositioner are measured as $\sim$100 and $\sim$\SI{2.3}{kHz} for the Z- and XY-displacements, respectively. The travel ranges are $\sim$6 and $\sim$\SI{34}{\micro m} for Z and XY, respectively. When this nanopositioner is used for hopping mode imaging of SICM with a $\sim$20-nm diameter tip, the vertical tip velocity can be increased to \SI{400}{nm/ms}; hence, the one-pixel acquisition time can be minimized to $\sim$\SI{1}{ms}. 

\end{abstract}


\maketitle 

\begin{figure}[!th]
\includegraphics{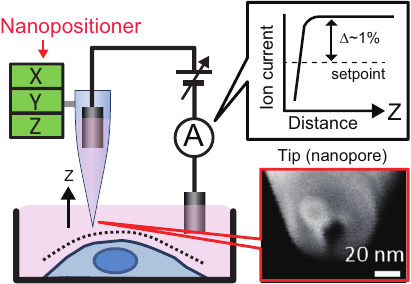}
\caption{
(Color online) Schematic of typical SICM setup showing working principle of SICM. The electrolyte-filled nanopipette mounted on the XYZ-nanopositioner has a nanopore at its distal end, as shown in the electron micrograph (right bottom). The ion current flowing through the nanopore induced by the application of bias voltage between two electrodes (one in the nanopipette, the other in the bath solution) is measured using the ion current detector. The ion current, which depends on the tip-surface separation as illustrated (right top), is used to control the tip Z-position during the XY-scanning of the nanopipette over the sample surface.
}
\label{f1}
\end{figure}

In biological research, atomic force microcopy (AFM)~\cite{binnig1986atomic} has been widely used to visualize the topographic structure of biological specimens under physiological liquid environments~\cite{dufrene2017imaging}. Nevertheless, the cantilever tip must make contact with the sample in the solution, and even exerted forces below 10 pN considerably deform extremely soft surfaces~\cite{zhou2013potentiometric,ushiki2012scanning} such as the plasma membranes of live eukaryotic cells, which prohibits high-resolution surface imaging~\cite{seifert2015comparison}. Scanning ion conductance microscopy (SICM) is an alternative imaging method based on an entirely different working principle for capturing a topographic image~\cite{hansma1989scanning} (Fig. \ref{f1}).
SICM uses an electrolyte-filled glass pipette (nanopipette) as a probe and relies on an ion current flowing between an electrode inside the nanopipette and another in an external bath solution. The ion current passing through the opening of the nanopipette is sensitive to the tip-sample surface separation~\cite{korchev1997scanning,novak2014imaging}; therefore, SICM can capture topographic images without any tip-sample contact.

However, the temporal resolution of SICM is much lower than that of AFM, especially when SICM is operated in the hopping mode~\cite{novak2009nanoscale}, where the tip is moved up and down to avoid lateral tip-sample contact. This low temporal resolution, due to the time delay in the vertical tip-position control, limits the fall velocity of the tip, $v_\textrm{f}$; namely, the speed at which the tip approaches the sample surface. Since glass nanopipettes are very fragile, they are easily damaged when they contact surfaces. To avoid surface contact, $v_\textrm{f}$ must be restricted. When the bandwidth of the ion current detection is sufficiently high, the low mechanical resonance frequency of the Z-nanopositioner is the dominant contributor to the delay~\cite{yong2010reducing}. The difficulty to decrease the delay is how to assemble the nanopipette to the Z-nanopositioner since the nanopipette is very massive and large compared to AFM cantilever tip. For instance, commercial SICM systems use a long-travel-range ($\sim$\SI{25}{\micro m}) Z-nanopositioner whose resonance frequency is less than 1 kHz when the nanopipette is assembled, producing a long delay (longer than 1 ms) for vertical tip positioning. As a result, it typically takes 10--100 ms for one-pixel acquisition when imaging samples showing very rough surfaces~\cite{novak2009nanoscale}. Thus, in SICM, it is difficult to visualize morphological changes in samples that occur in under a minute, which significantly limits the applicability of SICM in biological studies.

\begin{figure*}[!t]
\includegraphics{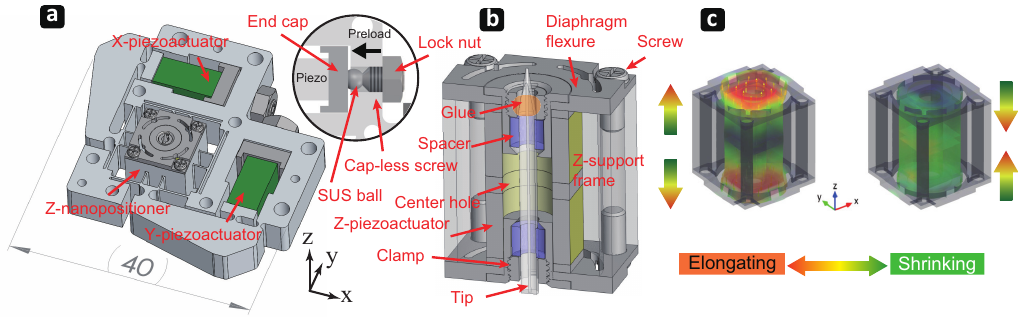}
\caption{
(Color online) (a) Drawing showing structure of XYZ-nanopositioner designed in this study. The base frame is \SI{5}{mm} thick and \SI{40}{mm}  wide. The lateral scan is driven with identical piezoactuators via mechanical amplification. Preloads are applied to the X- and Y-piezoactuators using the end caps, each of which is pushed with the capless screw via a stainless steel ball. To stabilize the preload, each screw is fixed with a lock nut. (b) Cross-sectional view of the Z-nanopositioner. The glue is only applied to a contact area between the top clamp and the nanopipette. (c) Displacement of Z-nanopositioner in the elongation (left) and shrinkage (right) actuations, as simulated by FEA. The arrows represent the vectors of local displacement from its original position.
}
\label{f2}
\end{figure*}


In this letter, we describe the design of a tip-scan-type high-speed XYZ-nanopositioner for SICM. One of the distinctive features of the developed nanopositioner is the large product of ``travel range $\times$ resonance frequency'' for the Z-positioner, which represents the positioner performance because the travel range and resonance frequency are tradeoffs. The product of our Z-positioner, \SI{6}{\micro m} $\times$ \SI{100}{kHz} , is more than ten-fold larger than those of conventional SICM systems and 2--3 times larger than that of the Z-scanner employed in high-speed AFM~\cite{ando2008high}. The performance can significantly increase $v_\textrm{f}$ to 400--\SI{600}{nm/ms} and hence minimize the one-pixel acquisition time to $\sim$\SI{1}{ms}. Here, we show the detailed design and performance of our nanopositioner and demonstrate the stable high-speed imaging of test samples at \SI{3.5}{s} per frame over a scan area of $\sim$25 $\times$ \SI{25}{\micro m^2} with 50 $\times$ 50 pixels.

The Z-nanopositioner mounted on the XY-nanopositioner, as illustrated in Fig. \ref{f2}(a), is designed to achieve a high mechanical resonance frequency and hence a large $v_\textrm{f}$. The frame body of the XYZ-nanopositioner was fabricated from alloy A7075. Before constructing the actual device, we conducted a finite-element analysis (FEA) to estimate the mechanical properties of designed structures, using a commercially available finite-element package, COMSOL Multiphysics 5.0 (COMSOL AB). The following mechanical parameters for the alloy A7075/piezoactuator were used in the FEA simulation: Young's modulus, 72/\SI{33.9}{GPa}; Poisson's ratio, 0.3/0.3; density, 2810/\SI{7800}{kg/m^3}. The central idea for suppressing unwanted mechanical vibrations in the Z-nanopositioner is the use of momentum cancellation~\cite{ando2001high}. That is, the hollow Z-piezoactuator (AE0505D08D-H0F, NEC/Tokin) is sandwiched with a pair of identical diaphragm-like flexures~\cite{yong2012design}[Fig. \ref{f2}(b)] so that the center of mass of the Z-piezoactuator shows negligible change during fast displacement. The flexures were designed to have a stiffness of $\sim$\SI{18.2}{N/\micro m}, which provides a suitable preload to the actuator ($\sim$10--\SI{20}{N}). The magnitude of the preload is adjustable with screws.

\begin{figure}[!t]
\includegraphics{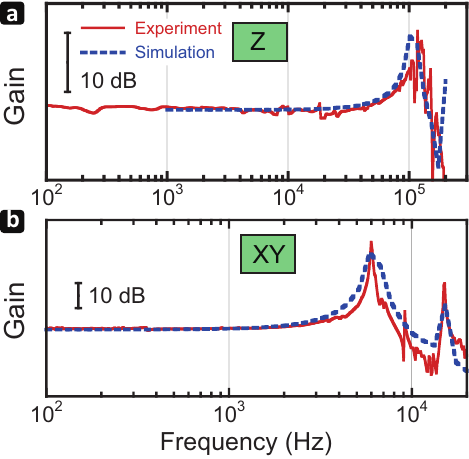}
\caption{
(Color online) Transfer function of nanopositioner for (a) vertical and (b) lateral displacements (phase is not shown). The solid and broken lines indicate the results from experiments and FEA simulations, respectively. The transfer functions were obtained for the 300-\si{\micro m}-thick beam flexures. The resonance frequency for the lateral displacement decreases from $\sim$6 to \SI{4.5}{kHz} when the full components of Z-positioner are assembled with the XY-positioner.
}
\label{f3}
\end{figure}

This design can mechanically cancel dynamic forces exerted onto the Z-support frame when the Z-piezoactuator is quickly displaced. Moreover, it permits the Z-positioner to have a resonance frequency comparable to that of the Z-piezoactuator under free oscillation. Injecting epoxy glue between the diaphragm flexures and the Z-piezoactuator improved the mechanical stability of the Z-positioner and enabled its robust long-term actuation. The $\sim$12--14-mm-long nanopipette is mechanically connected to the top flexure by gluing the nanopipette only to the top clamp [Fig. \ref{f2}(b)] so that the nanopipette can move with the z-movement of the top flexure. In order to avoid unintentional mechanical vibrations of the nanopipette, the nanopipette length was shortened as much as possible and the nanopipette was supported by the spacers made of an elastic material at the top and bottom clamps. A polyimide tube inserted between the nanopipette at the glue point enhances the fixing strength between the nanopipette to the clamp. This assembly only decreases the resonance frequency of the Z-piezoactuator by 10\% since the total mass of the nanopipette and clamp are much smaller than that of the Z-piezoactuator. Importantly, undesirable vibrations were suppressed in the Z-positioner below the first resonance frequency at \SI{100}{kHz}  [Fig. \ref{f3}(a)]. The measured first resonance frequency agreed reasonably well with the result of the FEA simulation [Fig. \ref{f3}(b)]. The travel range of Z-nanopositioner was $\sim$\SI{6}{\micro m}, as measured with a laser vibrometer (NLV-2500, Polytech), approximately half the original maximum displacement of the Z-piezoactuator, as expected. Since the travel range is inversely proportional to the resonance frequency~\cite{yong2012invited}, the performance of our Z-nanopositioner can be evaluated from the product of travel range $\times$ resonance frequency, which at \SI{6}{\micro m} $\times$ \SI{100}{kHz}  significantly exceeds (by more than 10-fold) the \SI{25}{\micro m} $\times$ 1--2 kHz shown in conventional SICM nanopositioners.

\begin{figure}[!t]
\includegraphics{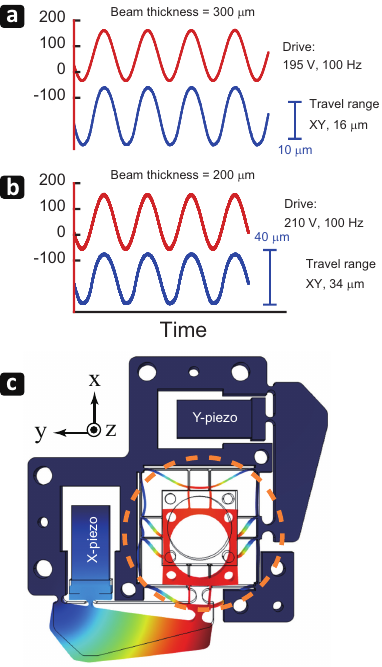}
\caption{
(Color online) Lateral displacements of the nanopositioner under 100-Hz sinusoidal voltage application. The beam flexures are (a) 300 and (b) \SI{200}{\micro m} thick. (c) FEA simulation result for the elongation of the X-piezoactuator. In this simulation, the positions of the backside of the piezoactuator and the screw holes in the scanner frame were fixed.
}
\label{f4}
\end{figure}

For XY-displacements, we used mechanical amplification ~\cite{yong2009design} to magnify the original travel range (\SI{9}{\micro m} at 150 V) of XY-piezoactuators (AE0505D08DF, NEC/Tokin). In our XY-nanopositioner design, the XY-travel range can be controlled by varying the thickness of the beam flexures that connect the Z-nanopositioner and the surrounding base frame, indicated by the broken circle in [Fig. \ref{f4}(c)], without changing the overall dimensions of the nanopositioner. The measured travel range increased from 16 [Fig. \ref{f4}(a)] to \SI{34}{\micro m} [Fig. \ref{f4}(b)] with decreasing beam-flexure thickness from 300 to \SI{200}{\micro m}, which is the practical limit for fabrication by wire electrical discharge machining. The mechanical amplification factors, 1.4 and 2.7, were consistent with the results of the FEA simulation (not shown). The measured dominant resonance frequency of the XY-nanopositioner assembled with the full components of the Z-nanopositioner decreased from 4.5 to \SI{2.3}{kHz}, adequate for achieving high-speed SICM, with decreasing beam-flexure thickness (not shown).

\begin{figure*}[!t]
\includegraphics{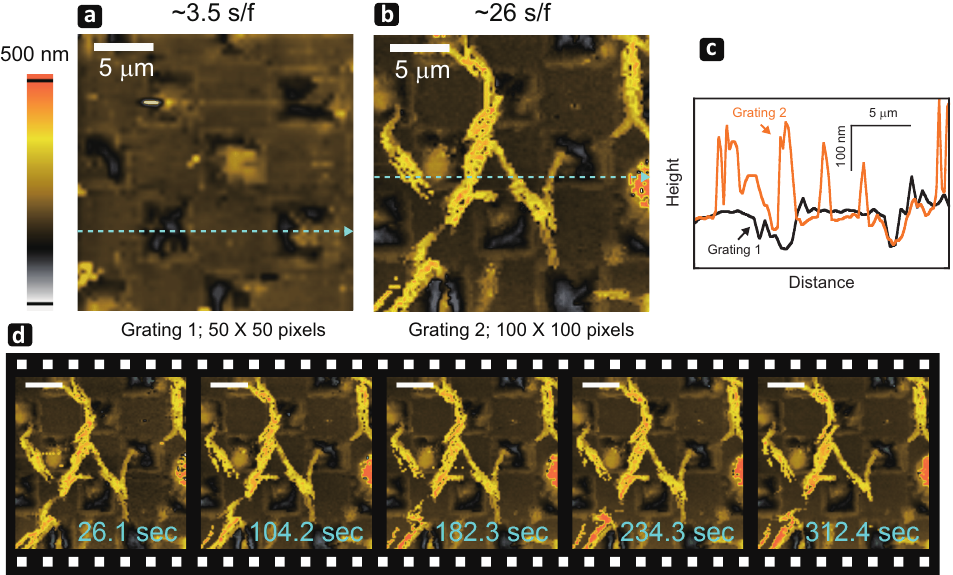}
\caption{
(Color online) SICM images of Gratings (a) 1 and (b) 2 captured at 3.5 and \SI{26}{s/frame}, respectively. (c) Height profiles of images (a) and (b) along the light blue broken lines shown in (a) and (b). (d) Five images of Grating 2 clipped from successive images captured at $\sim$\SI{26}{s/frame} for longer than \SI{5}{min}. Scale bars, \SI{5}{\micro m}.
}
\label{f5}
\end{figure*}

We first captured hopping-mode SICM images of Grating 1 samples immersed in a solution containing \SI{0.15}{M} KCl. The $v_\textrm{f}$ and hopping amplitude were set at \SI{400}{nm/ms} and 300 nm, respectively. Figure \ref{f5}(a) shows a topographic image captured at \SI{3.5}{s/frame} for a scan range of $\sim$25 $\times$ \SI{25}{\micro m^2} with 50 $\times$ 50 pixels. The image shows a grating pitch with variations of 20\% in the X- and Y-directions, mainly due to the hysteresis effect of the XY-piezoactuators and partially due to the cross-coupling between them. Since it is possible to compensate for these effects by exploiting previously reported methods~\cite{yong2010reducing,watanabe2013wide}, we do not focus on this issue in this study. In Fig. \ref{f5}(a), no notable undesirable vibrations appeared even at the left edge region of the image, where the scanning direction was inverted and hence the X-positioner was driven at high frequencies. The $v_\textrm{f}$ value, \SI{400}{nm/ms}, used in this imaging is significantly larger than the typical $v_\textrm{f}$ values used in conventional SICM systems ($\sim$20 nm/ms). Although the comparable $v_\textrm{f}$ value $\sim$\SI{500}{nm/ms} was reported for a large nanopore diameter $\sim$ 100 nm~\cite{novak2014imaging}, attainable $v_\textrm{f}$ value will decrease to $\sim$\SI{100}{nm/ms} when the previous report uses an identical nanopore diameter ($\sim$20 nm) used in this study. Therefore, our positioner significantly improved the $v_\textrm{f}$.

We first captured hopping-mode SICM images of Grating 1 samples immersed in a solution containing \SI{0.15}{M} KCl. The $v_\textrm{f}$ and hopping amplitude were set at \SI{400}{nm/ms} and 300 nm, respectively. Figure \ref{f5}(a) shows a topographic image captured at 3.5 s/frame for a scan range of $\sim$25 $\times$ \SI{25}{\micro m^2} with 50 $\times$ 50 pixels. The image shows a grating pitch with variations of 20\% in the X- and Y-directions, mainly due to the hysteresis effect of the XY-piezoactuators and partially due to the cross-coupling between them. Since it is possible to compensate for these effects by exploiting previously reported methods~\cite{yong2010reducing,watanabe2013wide}, we do not focus on this issue in this study. In Fig. \ref{f5}(a), no notable undesirable vibrations appeared even at the left edge region of the image, where the scanning direction was inverted and hence the X-positioner was driven at high frequencies. The $v_\textrm{f}$ value, \SI{400}{nm/ms}, used in this imaging is significantly larger than the typical $v_\textrm{f}$ values used in conventional SICM systems ($\sim$\SI{20}{nm/ms}). Although the comparable $v_\textrm{f}$ value $\sim$\SI{500}{nm/ms} was reported for a large nanopore diameter $\sim$ 100 nm~\cite{novak2014imaging}, atainable $v_\textrm{f}$ value will decrease to $\sim$\SI{100}{nm/ms} when the previous report uses an identical nanopore diameter ($\sim$\SI{20}{nm}) used in this study. Therefore, our positioner significantly improved the $v_\textrm{f}$.

To further evaluate the performance of our XYZ-nanopositioner, we captured a topographic image of Grating 2 with rougher surfaces than Grating 1, in which wire-like objects were seen on the surface, as shown in Fig. \ref{f5}(b). This image was captured at $\sim$26 s/frame for a scan range of $\sim$25 $\times$ \SI{25}{\micro m^2} with 100 $\times$ 100 pixels, using a hopping amplitude of \SI{600}{nm}and $v_\textrm{f}$ of \SI{400}{nm/ms}. By increasing the hopping amplitude to \SI{600}{nm}, wire-like objects likely to be partially suspended between grids [Fig. \ref{f5}(c)] were captured without their disruption, although the time required for one-pixel acquisition increased from $\sim$750 to \SI{1500}{\micro s}. Figure \ref{f5}(d) shows successive images of Grating 2 captured for longer than \SI{5}{min}. Since touching the surface manifests as `tail like patterns' in observed images regardless with the stiffness of wire-like objects~\cite{klenerman2011imaging}, no noticeable changes in these images demonstrate the robustness of our XYZ-positioning.

In summary, we developed a tip-scan-type high-speed XYZ-nanopositioner for SICM. The installation of the nanopipette mount in the center of a hollow piezoactuator and the use of a momentum cancellation mechanism for the Z-positioner resulted in a high resonance frequency of \SI{100}{kHz}  even with a relatively long travel range, \SI{6}{\micro m}. The improved time delay of the Z-nanopositioner increases the $v_\textrm{f}$ by more than 20 times that of conventional SICM nanopositioners. These excellent features allowed the high-speed imaging of Grating 1 at 3.5 s/frame for a scan area of 25 $\times$ \SI{25}{\micro m^2} with 50 $\times$ 50 pixels without generating undesirable vibrations. Moreover, a rougher surface could be captured without decreasing $v_\textrm{f}$. This study is the first step toward achieving high-speed SICM that can capture biological samples in dynamic action in real time.


This work was supported by the grant for `JST-SENTAN' (to S.W.) and the Grant for Young Scientists from Hokuriku Bank (to S.W.) and JSPS KAKENHI; Grant Numbers JP26790048 (to S.W.), JP16H00799 (to S.W.), and JP26119003 (to T.A.).


\bibliography{../../../Biblio/reference_20141021.bib}

\end{document}